\documentclass[11pt,a4paper]{article}
\usepackage[hyperref]{acl2018}
\usepackage{times}
\usepackage{booktabs}
\usepackage{graphicx}
\usepackage{subcaption}
\usepackage{caption} 
\usepackage{url}

\aclfinalcopy 

\title{Music Genre Classification using Machine Learning Techniques}

\author{Hareesh Bahuleyan \\
  University of Waterloo, ON, Canada \\  
  {\tt hpallika@uwaterloo.ca}}

\date{}
\begin{document}
\maketitle
\begin{abstract}
Categorizing music files according to their genre is a challenging task in the area of music information retrieval (MIR). In this study, we compare the performance of two classes of models. The first is a deep learning approach wherein a CNN model is trained end-to-end, to predict the genre label of an audio signal, solely using its spectrogram. The second approach utilizes hand-crafted features, both from the time domain and frequency domain. We train four traditional machine learning classifiers with these features and compare their performance. The features that contribute the most towards this classification task are identified. The experiments are conducted on the \textit{Audio set} data set and we report an AUC value of $0.894$ for an ensemble classifier which combines the two proposed approaches.\footnote{The code has been opensourced and is available at \url{https://github.com/HareeshBahuleyan/music-genre-classification}}
\end{abstract}

\section{Introduction}
With the growth of online music databases and easy access to music content, people find it increasing hard to manage the songs that they listen to. One way to categorize and organize songs is based on the genre, which is identified by some characteristics of the music such as rhythmic structure, harmonic content and instrumentation \cite{tzanetakis2002musical}. Being able to automatically classify and provide tags to the music present in a user's library, based on genre, would be beneficial for audio streaming services such as Spotify and iTunes. This study explores the application of machine learning (ML) algorithms to identify and classify the genre of a given audio file. The first model described in this paper uses convolutional neural networks \cite{krizhevsky2012imagenet}, which is trained end-to-end on the MEL spectrogram of the audio signal. In the second part of the study, we extract features both in the time domain and the frequency domain of the audio signal. These features are then fed to conventional machine learning models namely Logistic Regression, Random Forests \cite{breiman2001random}, Gradient Boosting \cite{friedman2001greedy} and Support Vector Machines which are trained to classify the given audio file. The models are evaluated on the \textit{Audio Set} dataset \cite{gemmeke2017audio}. We compare the proposed models and also study the relative importance of different features. 

The rest of this paper is organized as follows. Section \ref{sec:litrev} describes the existing methods in the literature for the task of music genre classification. Section \ref{sec:dataset} is an overview of the the dataset used in this study and how it was obtained. The proposed models and the implementation details are discussed in Section \ref{sec:method}. The results are reported in Section \ref{sec:results}, followed by the conclusions from this study in Section \ref{sec:conclusion}.

\section{Literature Review}
\label{sec:litrev}
Music genre classification has been a widely studied area of research since the early days of the Internet. \newcite{tzanetakis2002musical} addressed this problem with supervised machine learning approaches such as Gaussian Mixture model and $k$-nearest neighbour classifiers. They introduced 3 sets of features for this task categorized as timbral structure, rhythmic content and pitch content. Hidden Markov Models (HMMs), which have been extensively used for speech recognition tasks, have also been explored for music genre classification \cite{scaringella2005modeling,soltau1998recognition}. Support vector machines (SVMs) with different distance metrics are studied and compared in \newcite{mandel2005song} for classifying genre.  

In \newcite{lidy2005evaluation}, the authors discuss the contribution of psycho-acoustic features for recognizing music genre, especially the importance of STFT taken on the Bark Scale \cite{zwicker1999psychoacoustics}. Mel-frequency cepstral coefficients (MFCCs), spectral contrast and spectral roll-off were some of the features used by \cite{tzanetakis2002musical}. A combination of visual and acoustic features are used to train SVM and AdaBoost classifiers in \newcite{nanni2016combining}.  

With the recent success of deep neural networks, a number of studies apply these techniques to speech and other forms of audio data \cite{abdel2014convolutional,gemmeke2017audio}. Representing audio in the time domain for input to neural networks is not very straight-forward because of the high sampling rate of audio signals. However, it has been addressed in \newcite{van2016wavenet} for audio generation tasks. A common alternative representation is the spectrogram of a signal which captures both time and frequency information. Spectrograms can be considered as images and used to train convolutional neural networks (CNNs) \cite{wyse2017audio}. A CNN was developed to predict the music genre using the raw MFCC matrix as input in \newcite{li2010automatic}. In \newcite{lidy2016parallel}, a constant Q-transform (CQT) spectrogram was provided as input to the CNN to achieve the same task. 

This work aims to provide a comparative study between 1) the deep learning based models which only require the spectrogram as input and, 2) the traditional machine learning classifiers that need to be trained with hand-crafted features. We also investigate the relative importance of different features.

\section{Dataset}
\label{sec:dataset}
In this work, we make use of \textit{Audio Set}, which is a large-scale human annotated database of sounds \cite{gemmeke2017audio}. The dataset was created by extracting \textbf{10-second sound clips} from a total of 2.1 million YouTube videos. The audio files have been annotated on the basis of an ontology which covers 527 classes of sounds including musical instruments, speech, vehicle sounds, animal sounds and so on\footnote{\url{https://research.google.com/audioset/ontology/index.html}}. This study requires only the audio files that belong to the music category, specifically having one of the seven genre tags shown in Table \ref{tab:data}.

\begin{table}[htbp]
  \centering
  \caption{Number of instances in each genre class}
    \begin{tabular}{|r|c|c|}
    \toprule
          & \textbf{Genre} & \textbf{Count} \\
    \midrule
    \multicolumn{1}{|c|}{1} & Pop Music & 8100 \\
    \midrule
    \multicolumn{1}{|c|}{2} & Rock Music & 7990 \\
    \midrule
    \multicolumn{1}{|c|}{3} & Hip Hop Music & 6958 \\
    \midrule
    \multicolumn{1}{|c|}{4} & Techno & 6885 \\
    \midrule
    \multicolumn{1}{|c|}{5} & Rhythm Blues & 4247 \\
    \midrule
    \multicolumn{1}{|c|}{6} & Vocal & 3363 \\
    \midrule
    \multicolumn{1}{|c|}{7} & Reggae Music & 2997 \\
    \midrule
          & \textbf{Total} & \textbf{40540} \\
    \bottomrule
    \end{tabular}%
  \label{tab:data}%
\end{table}%

The number of audio clips in each category has also been tabulated. The raw audio clips of these sounds have not been provided in the \textit{Audio Set} data release. However, the data provides the \texttt{YouTubeID} of the corresponding videos, along with the start and end times. Hence, the first task is to retrieve these audio files. For the purpose of audio retrieval from YouTube, the following steps were carried out:
\begin{enumerate}
\item A command line program called \texttt{youtube-dl} \cite{gonzalez2006youtube} was utilized to download the video in the \texttt{mp4} format.
\item The \texttt{mp4} files are converted into the desired \texttt{wav} format using an audio converter named \texttt{ffmpeg} \cite{tomar2006converting} (command line tool).
\end{enumerate}
Each \texttt{wav} file is about 880 KB in size, which means that the total data used in this study is approximately 34 GB. 

\begin{figure*}[!t]
	\centering
	\includegraphics[width=1.0\textwidth]{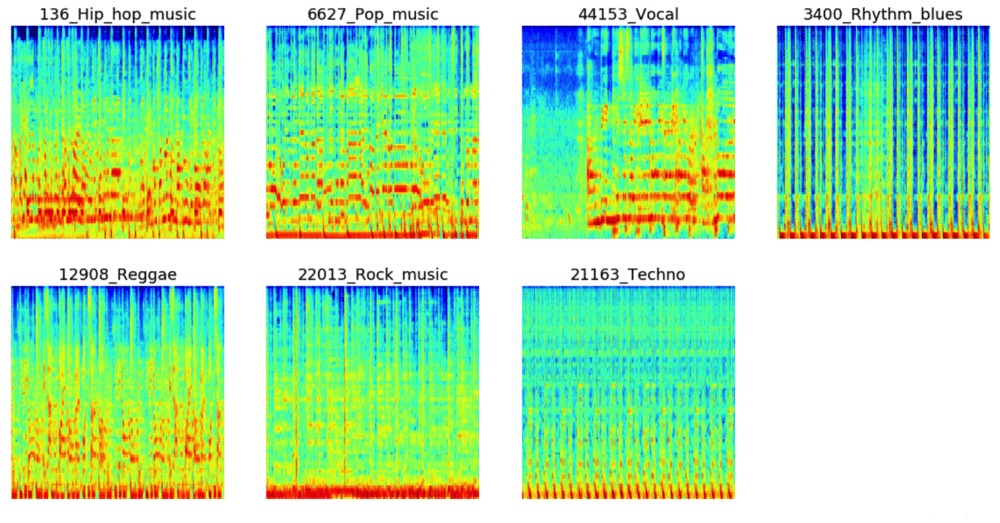}
    \caption{Sample spectrograms for 1 audio signal from each music genre}
    \label{fig:spectrograms}%
\end{figure*}

\begin{figure*}[th]
	\centering
	\includegraphics[width=1.0\textwidth]{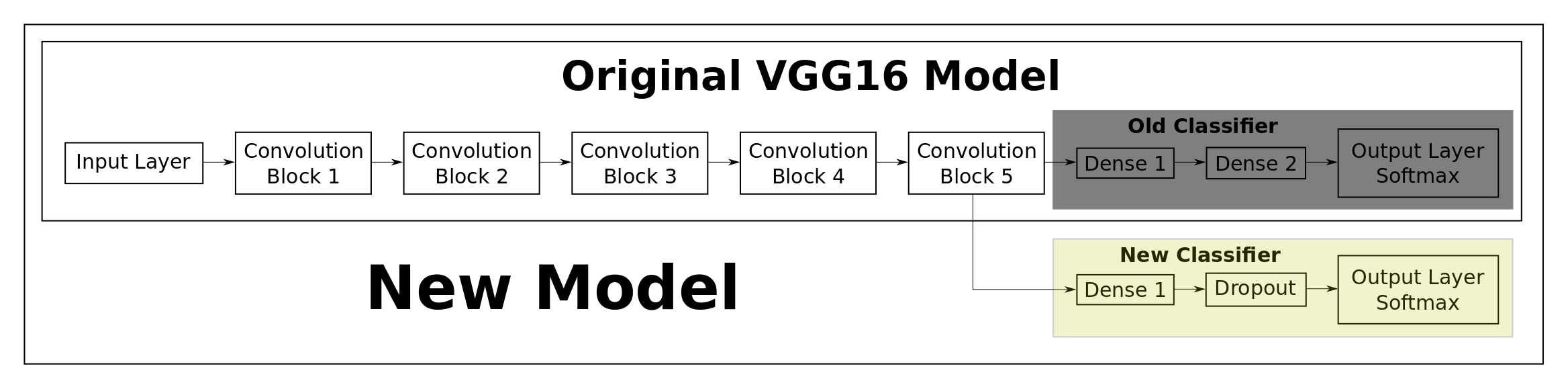}
    \caption{Convolutional neural network architecture (Image Source: \href{https://github.com/Hvass-Labs/TensorFlow-Tutorials/blob/master/10\_Fine-Tuning.ipynb}{Hvass Tensorflow Tutorials})}
    \label{fig:architecture}%
\end{figure*}

\section{Methodology}
\label{sec:method}
This section provides the details of the data pre-processing steps followed by the description of the two proposed approaches to this classification problem. 

\subsection{Data Pre-processing}
In order to improve the Signal-to-Noise Ratio (SNR) of the signal, a pre-emphasis filter, given by Equation \ref{eqn:pre-emp} is applied to the original audio signal. 
\begin{equation}
y(t) = x(t) - \alpha*x(t-1)
\label{eqn:pre-emp}
\end{equation}
where, $x(t)$ refers to the original signal, and $y(t)$ refers to the filtered signal and $\alpha$ is set to 0.97. Such a pre-emphasis filter is useful to boost amplitudes at high frequencies \cite{kim2012power}. 

\subsection{Deep Neural Networks}
\label{sec:cnn}
Using deep learning, we can achieve the task of music genre classification without the need for hand-crafted features. Convolutional neural networks (CNNs) have been widely used for the task of image classification \cite{krizhevsky2012imagenet}. The 3-channel (RGB) matrix representation of an image is fed into a CNN which is trained to predict the image class. In this study, the sound wave can be represented as a spectrogram, which in turn can be treated as an image \cite{nanni2016combining}\cite{lidy2016parallel}. The task of the CNN is to use the spectrogram to predict the genre label (one of seven classes). 

\subsubsection{Spectrogram Generation}
\label{sec:specgen}
A spectrogram is a 2D representation of a signal, having time on the x-axis and frequency on the y-axis. A colormap is used to quantify the magnitude of a given frequency within a given time window. In this study, each audio signal was converted into a MEL spectrogram (having MEL frequency bins on the y-axis). The parameters used to generate the power spectrogram using STFT are listed below:
\begin{itemize}
\item Sampling rate (\texttt{sr}) = 22050
\item Frame/Window size (\texttt{n\_fft}) = 2048
\item Time advance between frames (\texttt{hop\_size}) = 512 (resulting in 75\% overlap)
\item Window Function: Hann Window
\item Frequency Scale: MEL
\item Number of MEL bins: 96
\item Highest Frequency (\texttt{f\_max}) = \texttt{sr/2}
\end{itemize}

\subsubsection{Convolutional Neural Networks}
From the Figure~\ref{fig:spectrograms}, one can understand that there exists some characteristic patterns in the spectrograms of the audio signals belonging to different classes. Hence, spectrograms can be considered as 'images' and provided as input to a CNN, which has shown good performance on image classification tasks. 
Each block in a CNN consists of the following operations\footnote{\url{https://ujjwalkarn.me/2016/08/11/intuitive-explanation-convnets/}}:
\begin{itemize}
\item \textbf{Convolution}: This step involves sliding a matrix filter (say 3x3 size) over the input image which is of dimension \texttt{image\_width x image\_height}. The filter is first placed on the image matrix and then we compute an element-wise multiplication between the filter and the overlapping portion of the image, followed by a summation to give a feature value. We use many such filters , the values of which are 'learned' during the training of the neural network via backpropagation.  
\item \textbf{Pooling}: This is a way to reduce the dimension of the feature map obtained from the convolution step, formally know as the process of \textit{down sampling}. For example, by max pooling with 2x2 window size, we only retain the element with the maximum value among the 4 elements of the feature map that are covered in this window. We keep moving this window across the feature map with a pre-defined stride. 
\item \textbf{Non-linear Activation}: The convolution operation is linear and in order to make the neural network more powerful, we need to introduce some non-linearity. For this purpose, we can apply an activation function such as ReLU\footnote{\url{https://en.wikipedia.org/wiki/Rectifier\_(neural\_networks)}} on each element of the feature map. 
\end{itemize}

\begin{figure*}[!t]
    \centering
    \begin{subfigure}[t]{0.45\textwidth}
\centering
	\includegraphics[width=\textwidth]{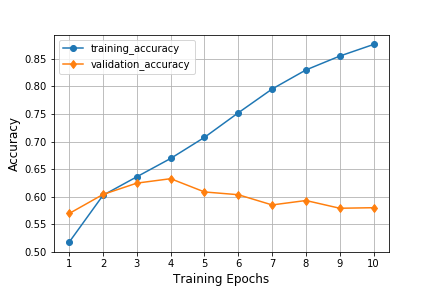}
    \caption{Accuracy}
    \label{fig:lc-acc}
    \end{subfigure}
    ~
    \begin{subfigure}[t]{0.45\textwidth}
	\centering
	\includegraphics[width=\textwidth]{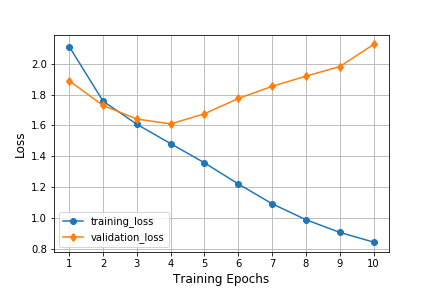}
    \caption{Loss}
    \label{fig:lc-loss}%
    \end{subfigure}%
    \caption{Learning Curves - used for model selection; Epoch 4 has the minimum validation loss and highest validation accuracy}
    \label{fig:lc}
\end{figure*}

In this study, a CNN architecture known as VGG-16, which was the top performing model in the ImageNet Challenge 2014 (classification + localization task) was used \cite{simonyan2014very}. The model consists of 5 convolutional blocks (conv base), followed by a set of densely connected layers, which outputs the probability that a given image belongs to each of the possible classes. 

For the task of music genre classification using spectrograms, we download the model architecture with pre-trained weights, and extract the conv base. The output of the conv base is then send to a new feed-forward neural network which in turn predicts the genre of the music, as depicted in Figure \ref{fig:architecture}. 

There are two possible settings while implementing the pre-trained model:
\begin{enumerate}
\item \textbf{Transfer learning}: The weights in the conv base are kept fixed but the weights in the feed-forward network (represented by the yellow box in Figure \ref{fig:architecture}) are allowed to be tuned to predict the correct genre label. 
\item \textbf{Fine tuning}: In this setting, we start with the pre-trained weights of VGG-16, but allow all the model weights to be tuned during training process. 
\end{enumerate}

The final layer of the neural network outputs the class probabilities (using the softmax activation function) for each of the seven possible class labels. Next, the cross-entropy loss is computed as follows:
\begin{equation}
\mathcal{L} = -\sum_{c=1}^{M} y_{o,c}*\log p_{o,c}
\end{equation}
where, $M$ is the number of classes; $y_{o,c}$ is a binary indicator whose value is 1 if observation $o$ belongs to class $c$ and $0$ otherwise; $p_{o,c}$ is the model's predicted probability that observation $o$ belongs to class $c$.
This loss is used to backpropagate the error, compute the gradients and thereby update the weights of the network. This iterative process continues until the loss converges to a minimum value. 

\subsubsection{Implementation Details}
\label{sec:imp}
The spectrogram images have a dimension of \texttt{216 x 216}. For the feed-forward network connected to the conv base, a 512-unit hidden layer is implemented. Over-fitting is a common issue in neural networks. In order to prevent this, two strategies are adopted:

\begin{enumerate}
\item \textbf{L2-Regularization} \cite{ng2004feature}: The loss function of the neural network is added with the term $\frac{1}{2}\lambda \sum_i{{w_i}^2}$, where $w$ refers to the weights in the neural networks. This method is used to penalize excessively high weights. We would like the weights to be diffused across all model parameters, and not just among a few parameters. Also, intuitively, smaller weights would correspond to a less complex model, thereby avoiding over-fitting. $\lambda$ is set to a value of $0.001$ in this study.
\item \textbf{Dropout} \cite{srivastava2014dropout}: This is a regularization mechanism in which we \textit{shut-off} some of the neurons (set their weights to zero) randomly during training. In each iteration, we thereby use a different combination of neurons to predict the final output. This makes the model generalize without any heavy dependence on a subset of the neurons. A dropout rate of $0.3$ is used, which means that a given weight is set to zero during an iteration, with a probability of $0.3$.
\end{enumerate}

The dataset is randomly split into train (90\%), validation (5\%) and test (5\%) sets. The same split is used for all experiments to ensure a fair comparison of the proposed models.

The neural networks are implemented in Python using Tensorflow \footnote{\url{http://tensorflow.org/}}; an NVIDIA Titan X GPU was utilized for faster processing. All models were trained for 10 epochs with a batch size of 32 with the ADAM optimizer \cite{kingma2014adam}. One epoch refers to one iteration over the entire training dataset.

Figure \ref{fig:lc} shows the learning curves - the loss (which is being optimized) keeps decreasing as the training progresses.  Although the training accuracy keeps increasing, the validation accuracy first increases and after a certain number of epochs, it starts to decrease. This shows the model's tendency to overfit on the training data. The model that is selected for evaluation purposes is the one that has the highest accuracy and lowest loss on the validation set (epoch 4 in Figure \ref{fig:lc}).

\subsubsection{Baseline Feed-forward Neural Network}
\label{sec:baseline}
To assess the performance improvement that can be achived by the CNNs, we also train a baseline feed-forward neural network that takes as input the same spectrogram image. The image which is a 2-dimensional vector of pixel values is unwrapped or flattened into a 1-dimensional vector. Using this vector, a simple 2-layer neural network is trained to predict the genre of the audio signal. The first hidden layer consists of 512 units and the second layer has 32 units, followed by the output layer. The activation function used is ReLU and the same regularization techniques described in Section \ref{sec:imp} are adopted.

\subsection{Manually Extracted Features}
\label{sec:feature}
In this section, we describe the second category of proposed models, namely the ones that require hand-crafted features to be fed into a machine learning classifier. Features can be broadly classified as time domain and frequency domain features. The feature extraction was done using \texttt{librosa}\footnote{\url{https://librosa.github.io/}}, a Python library.

\subsubsection{Time Domain Features}
These are features which were extracted from the raw audio signal.
\begin{enumerate}

\item \textbf{Central moments}: This consists of the mean, standard deviation, skewness and kurtosis of the amplitude of the signal. 

\item \textbf{Zero Crossing Rate (ZCR)}: A zero crosssing point refers to one where the signal changes sign from positive to negative \cite{gouyon2000use}. The entire 10 second signal is divided into smaller frames, and the number of zero-crossings present in each frame are determined. The frame length is chosen to be 2048 points with a hop size of 512 points. Note that these frame parameters have been used consistently across all features discussed in this section. Finally, the average and standard deviation of the ZCR across all frames are chosen as representative features. 
\item \textbf{Root Mean Square Energy (RMSE)}: The energy in a signal is calculated as:
\begin{equation}
\sum_{n=1}^{N} {|x(n)|}^2
\end{equation}
Further, the root mean square value can be computed as:
\begin{equation}
\sqrt{\frac{1}{N} \sum_{n=1}^{N} {|x(n)|}^2}
\end{equation}
RMSE is calculated frame by frame and then we take the average and standard deviation across all frames.

\item \textbf{Tempo}: In general terms, tempo refers to the how fast or slow a piece of music is; it is expressed in terms of Beats Per Minute (BPM). Intuitively, different kinds of music would have different tempos. Since the tempo of the audio piece can vary with time, we aggregate it by computing the mean across several frames. The functionality in \texttt{librosa} first computes a tempogram following \cite{grosche2010cyclic} and then estimates a single value for tempo.
\end{enumerate}

\subsubsection{Frequency Domain Features}
The audio signal can be transformed into the frequency domain by using the Fourier Transform. We then extract the following features. 
\begin{enumerate}
\item \textbf{Mel-Frequency Cepstral Coefficients (MFCC)}: Introduced in the early 1990s by Davis and Mermelstein, MFCCs have been very useful features for tasks such as speech recognition \cite{davis1990comparison}. First, the Short-Time Fourier-Transform (STFT) of the signal is taken with \texttt{n\_fft=2048} and \texttt{hop\_size=512} and a Hann window. Next, we compute the power spectrum and then apply the triangular MEL filter bank, which mimics the human perception of sound. This is followed by taking the discrete cosine transform of the logarithm of all filterbank energies, thereby obtaining the MFCCs. The parameter \texttt{n\_mels}, which corresponds to the number of filter banks, was set to 20 in this study.

\item \textbf{Chroma Features}: This is a vector which corresponds to the total energy of the signal in each of the 12 pitch classes. (C, C\#, D, D\#, E ,F, F\#, G, G\#, A, A\#, B) \cite{ellis2007chroma}. The chroma vectors are then aggregated across the frames to obtain a representative mean and standard deviation. 

\item \textbf{Spectral Centroid}:
For each frame, this corresponds to the frequency around which most of the energy is centered \cite{mir}.
It is a magnitude weighted frequency calculated as:
\begin{equation}
f_c = \frac{\sum_{k} S(k)f(k)}{\sum_{k} f{k}},
\end{equation}
where S(k) is the spectral magnitude of frequency bin k and  f(k) is the frequency corresponding to bin  k. 

\item \textbf{Spectral Band-width}:
The $p$-th order spectral band-width corresponds to the $p$-th order moment about the spectral centroid \cite{mir} and is calculated as  
\begin{equation}
[\sum_k (S(k)f(k) - f_c)^p]^\frac{1}{p}
\end{equation}
For example, $p = 2$ is analogous to a weighted standard deviation. 

\item \textbf{Spectral Contrast}: Each frame is divided into a pre-specified number of frequency bands. And, within each frequency band, the spectral contrast is calculated as the difference between the maximum and minimum magnitudes \cite{jiang2002music}.

\item \textbf{Spectral Roll-off}: This feature corresponds to the value of frequency below which 85\% (this threshold can be defined by the user) of the total energy in the spectrum lies \cite{mir}. 

\end{enumerate}

For each of the spectral features described above, the mean and standard deviation of the values taken across frames is considered as the representative final feature that is fed to the model.

The features described in this section would be would be used to train machine learning algorithms (refer Section \ref{sec:classifiers}).  The features that contribute the most in achieving a good classification performance will be identified and reported. 

\subsection{Classifiers}
\label{sec:classifiers}
This section provides a brief overview of the four machine learning classifiers adopted in this study.

\begin{enumerate}

\item \textbf{Logistic Regression (LR)}: This linear classifier is generally used for binary classification tasks. For this multi-class classification task, the LR is implemented as a one-vs-rest method. That is, 7 separate binary classifiers are trained. During test time, the class with the highest probability from among the 7 classifiers is chosen as the predicted class.

\item \textbf{Random Forest (RF)}: Random Forest is a ensemble learner that combines the prediction from a pre-specified number of decision trees. It works on the integration of two main principles: 1) each decision tree is trained with only a subset of the training samples which is known as bootstrap aggregation (or bagging) \cite{breiman1996bagging}, 2) each decision tree is required to make its prediction using only a random subset of the features \cite{amit1997shape}. The final predicted class of the RF is determined based on the majority vote from the individual classifiers.   

\item \textbf{Gradient Boosting (XGB)}: Boosting is another ensemble classifier that is obtained by combining a number of weak learners (such as decision trees). However, unlike RFs, boosting algorithms are trained in a sequential manner using forward stagewise additive modelling \cite{hastie2001elements}. 

During the early iterations, the decision trees learnt are fairly simple. As training progresses, the classifier become more powerful because it is made to focus on the instances where the previous learners made errors. At the end of training, the final prediction is a weighted linear combination of the output from the individual learners. XGB refers to eXtreme Gradient Boosting, which is an implementation of boosting that supports training the model in a fast and parallelized manner. 

\begin{table*}[!t]
  \centering
  \caption{Comparison of performance of the models on the test set}
    \begin{tabular}{l|c|c|c}
\cmidrule{2-4}    \multicolumn{1}{c}{} & \textbf{Accuracy} & \textbf{F-score} & \textbf{AUC} \\
    \midrule
    \multicolumn{1}{l}{\textbf{Spectrogram-based models}} & \multicolumn{1}{c}{} & \multicolumn{1}{c}{} &  \\
    \midrule
    VGG-16 CNN Transfer Learning & 0.63  & 0.61  & \textbf{0.891} \\
    VGG-16 CNN Fine Tuning & \textbf{0.64} & \textbf{0.61} & 0.889 \\
    Feed-forward NN baseline & 0.43  & 0.33  & 0.759 \\
    \midrule
    \multicolumn{1}{l}{\textbf{Feature Engineering based models}} & \multicolumn{1}{c}{} & \multicolumn{1}{c}{} &  \\
    \midrule
    Logistic Regression (LR) & 0.53  & 0.47  & 0.822 \\
    Random Forest (RF) & 0.54  & 0.48  & 0.840 \\
    Support Vector Machines (SVM) & 0.57  & 0.52  & 0.856 \\
    Extreme Gradient Boosting (XGB) & \textbf{0.59} & \textbf{0.55} & \textbf{0.865} \\
    \midrule
    \multicolumn{1}{l}{\textbf{Ensemble Classifiers}} & \multicolumn{1}{c}{} & \multicolumn{1}{c}{} &  \\
    \midrule
    VGG-16 CNN + XGB & \textbf{0.65} & \textbf{0.62} & \textbf{0.894} \\
    \bottomrule
    \end{tabular}%
  \label{tab:results}%
\end{table*}%

\item \textbf{Support Vector Machines (SVM)}: SVMs transform the original input data into a high dimensional space using a kernel trick \cite{cortes1995support}. The transformed data can be linearly separated using a hyperplane. The optimal hyperplane maximizes the margin. In this study, a radial basis function (RBF) kernel is used to train the SVM because such a kernel would be required to address this non-linear problem. Similar to the logistic regression setting discussed above, the SVM is also implemented as a one-vs-rest classification task. 

\end{enumerate}

\section{Evaluation}
\label{sec:eval}
\subsection{Metrics}
\label{sec:metrics}
In order to evaluate the performance of the models described in Section \ref{sec:method}, the following metrics will be used. 
\begin{itemize}
\item \textbf{Accuracy}: Refers to the percentage of correctly classified test samples. 
\item \textbf{F-score}: Based on the confusion matrix, it is possible to calculate the precision and recall. F-score\footnote{\url{https://en.wikipedia.org/wiki/F1\_score}} is then computed as the harmonic mean between precision and recall.
\item \textbf{AUC}: This evaluation criteria known as the area under the receiver operator characteristics (ROC) curve is a common way to judge the performance of a multi-class classification system. The ROC is a graph between the true positive rate and the false positive rate. A baseline model which randomly predicts each class label with equal probability would have an AUC of 0.5, and hence the system being designed is expected to have a AUC higher than 0.5.

\end{itemize}

\subsection{Results and Discussion}
\label{sec:results}
In this section, the different modelling approaches discussed in Section \ref{sec:method} are evaluated based on the metrics described in Section \ref{sec:metrics}. The values have been reported in Table \ref{tab:results}.

The best performance in terms of all metrics is observed for the convolutional neural network model based on VGG-16 that uses only the spectrogram to predict the music genre. It was expected that the fine tuning setting, which additionally allows the convolutional base to be trainable, would enhance the CNN model when compared to the transfer learning setting. However, as shown in Table \ref{tab:results}, the experimental results show that there is no significant difference between transfer learning and fine-tuning. The baseline feed-forward neural network that uses the unrolled pixel values from the spectrogram performs poorly on the test set. This shows that CNNs can significantly improve the scores on such an image classification task. 

Among the models that use manually crafted features, the one with the least performance is the Logistic regression model. This is expected since logistic regression is a linear classifier. SVMs outperform random forests in terms of accuracy. However, the XGB version of the gradient boosting algorithm performs the best among the feature engineered methods. 

\begin{figure*}[!t]
	\centering
	\includegraphics[width=.8\textwidth]{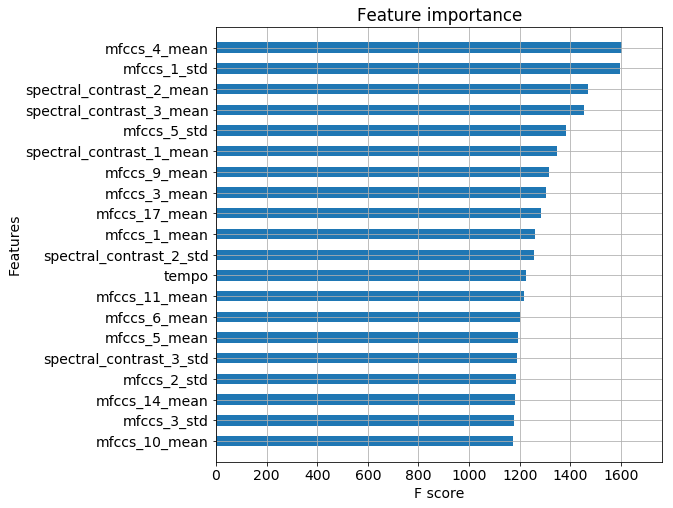}
    \caption{Relative importance of features in the XGBoost model; the top 20 most contributing features are displayed}
    \label{fig:feature-imp}%
\end{figure*}

\subsubsection{Most Important Features}
In this section, we investigate which features contribute the most during prediction, in this classification task. To carry out this experiment, we chose the XGB model, based on the results discussed in the previous section. To do this, we rank the top 20 most useful features based on a scoring metric (Figure \ref{fig:feature-imp}). The metric is calculated as the number of times a given feature is used as a decision node among the individual decision trees that form the gradient boosting predictor. 

As can be observed from Figure \ref{fig:feature-imp}, Mel-Frequency Cepstral Coefficients (MFCC) appear the most among the important features. Previous studies have reported MFCCs to improve the performance of speech recognition systems \cite{ittichaichareon2012speech}. Our experiments show that MFCCs contribute significantly to this task of music genre classification. The mean and standard deviation of the spectral contrasts at different frequency bands are also important features. The music tempo, calculated in terms of beats per minute also appear in the top 20 useful features.

Next, we study how much of performance in terms of AUC and accuracy, can be obtained by just using the top $N$ while training the model. From Table \ref{tab:topn} it can be seen that with only the top 10 features, the model performance is surprisingly good. In comparison to the full model which has 97 features, the model with the top 30 features has only a marginally lower performance (2 points on the AUC metric and 4 point on the accuracy metric).  

\begin{table}[htbp]
  \centering
  \caption{Ablation Study: Comparing XGB performance keeping only top $N$ features}
    \begin{tabular}{|c|c|c|}
    \toprule
    \textbf{N}     & \textbf{AUC}   & \textbf{Accuracy} \\
    \midrule
    10    & 0.803 & 0.47 \\
    20    & 0.837 & 0.52 \\
    30    & 0.845 & 0.55 \\
    97    & 0.865 & 0.59 \\
    \bottomrule
    \end{tabular}%
  \label{tab:topn}%
\end{table}%

The final experiment in this section is comparison of time domain and frequency domain features listed in Section \ref{sec:feature}. Two XGB models were trained - one with only time domain features and the other with only frequency domain features. Table \ref{tab:timefreq} compares the results in terms of AUC and accuracy. This experiment further confirms the fact that frequency domain features are definitely better than time domain features when it comes to modelling audio for machine learning tasks.  

\begin{table}[htbp]
  \centering
  \caption{Comparison of Time Domain features and Frequency Domain features}
    \begin{tabular}{|c|c|c|}
    \toprule
    \textbf{Model}     & \textbf{AUC}   & \textbf{Accuracy} \\
    \midrule
    Time Domain only    & 0.731 & 0.40 \\
    Frequency Domain only     & 0.857 & 0.57 \\
    Both    & 0.865 & 0.59 \\
    \bottomrule
    \end{tabular}%
  \label{tab:timefreq}%
\end{table}%

\begin{figure*}[t!]
    \centering
    \begin{subfigure}[t]{0.48\textwidth}
        \centering
        \includegraphics[width=1.0\textwidth]{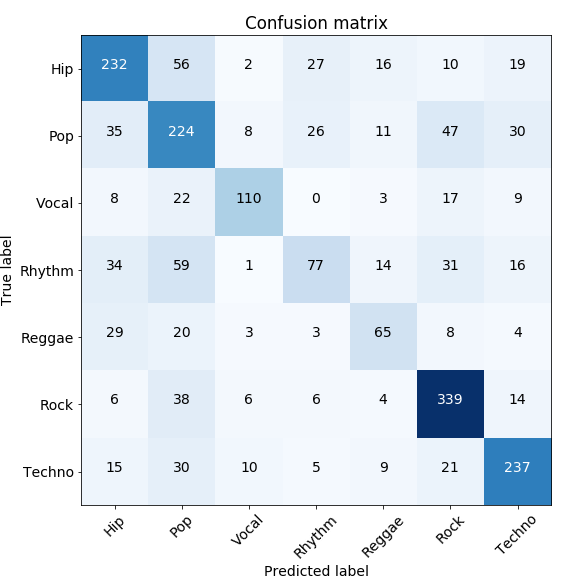}
        \caption{VGG-16 CNN Transfer Learning}
    \end{subfigure}
    ~
    \begin{subfigure}[t]{0.48\textwidth}
        \centering
        \includegraphics[width=1.0\textwidth]{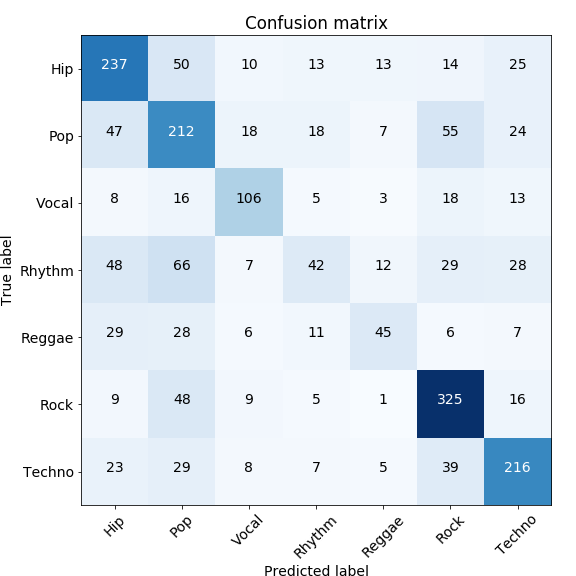}
        \caption{Extreme Gradient Boosting}
    \end{subfigure}%
    ~
    \begin{subfigure}[t]{0.48\textwidth}
        \centering
        \includegraphics[width=1.0\textwidth]{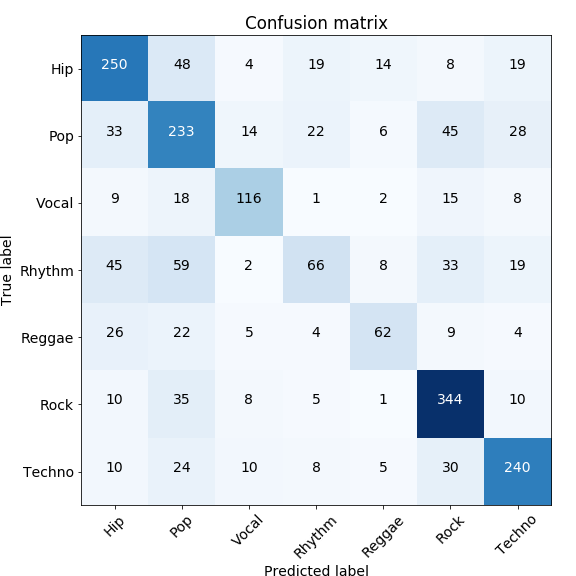}
        \caption{Ensemble Model}
    \end{subfigure}%
    \caption{Confusion Matrices of the best performing models}
\label{fig:cm}
\end{figure*}

\subsubsection{Confusion Matrix}
Confusion matrix is a tabular representation which enables us to further understand the strengths and weaknesses of our model. Element $a_{ij}$ in the matrix refers to the number of test instances of class $i$ that the model predicted as class $j$. Diagonal elements $a_{ii}$ corresponds to the correct predictions. Figure \ref{fig:cm} compares the confusion matrices of the best performing CNN model and XGB, the best model among the feature-engineered classifiers. Both models seems to be good at predicting the class 'Rock' music. However, many instances of class 'Hip Hop' are often confused with class 'Pop' and vice-versa. Such a behaviour is expected when the genres of music are very close. Some songs may fall into multiple genres, even as much that it may be difficult for humans to recognize the exact genre. 

\subsubsection{Ensemble Classifier}
Ensembling is a commonly adopted practice in machine learning, wherein, the results from different classifiers are combined. This is done by either majority voting or by averaging scores/probabilities. Such an ensembling scheme which combines the prediction powers of different classifiers makes the overall system more robust. In our case, each classifier outputs a prediction probability for each of the class labels. Hence, averaging the predicted probabilities from the different classifiers would be a straight-forward way to do ensemble learning.

\begin{figure}[th]
	\centering
	\includegraphics[width=1.0\linewidth]{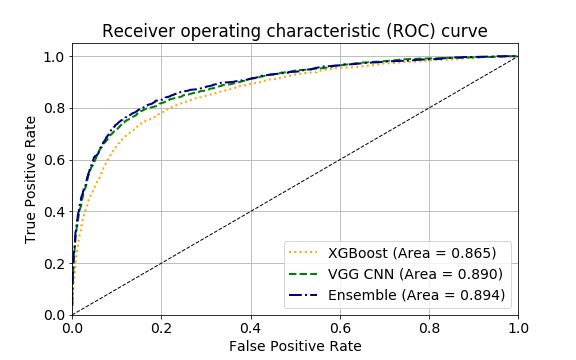}
    \caption{ROC Curves for the best performing models and their ensemble}
    \label{fig:roc-curves}%
\end{figure}

The methodologies described in \ref{sec:cnn} and \ref{sec:classifiers} use very different sources of input, the spectrograms and the hand-crafted features respectively. Hence, it makes sense to combine the models via ensembling. In this study, the best CNN model namely, VGG-16 Transfer Learning is ensembled with XGBoost the best feature engineered model by averaging the predicted probabilities. As shown in Table \ref{tab:results}, this ensembling is beneficial and is observed to outperform the all individual classifiers. The ROC curve for the ensemble model is above that of VGG-16 Fine Tuning and XGBoost as illustrated in Figure \ref{fig:roc-curves}. 

\section{Conclusion}
\label{sec:conclusion}
In this work, the task of music genre classification is studied using the Audioset data. We propose two different approaches to solving this problem. The first involves generating a spectrogram of the audio signal and treating it as an image. An CNN based image classifier, namely VGG-16 is trained on these images to predict the music genre solely based on this spectrogram. The second approach consists of extracting time domain and frequency domain features from the audio signals, followed by training traditional machine learning classifiers based on these features. XGBoost was determined to be the best feature-based classifier; the most important features were also reported. The CNN based deep learning models were shown to outperform the feature-engineered models. We also show that ensembling the CNN and XGBoost model proved to be beneficial. It is to be noted that the dataset used in this study was audio clips from YouTube videos, which are in general very noisy. Futures studies can identify ways to pre-process this noisy data before feeding it into a machine learning model, in order to achieve better performance.  

\bibliography{acl2018}

\begin{thebibliography}{}
\expandafter\ifx\csname natexlab\endcsname\relax\def\natexlab#1{#1}\fi

\bibitem[{Abdel-Hamid et~al.(2014)Abdel-Hamid, Mohamed, Jiang, Deng, Penn, and
  Yu}]{abdel2014convolutional}
Ossama Abdel-Hamid, Abdel-rahman Mohamed, Hui Jiang, Li~Deng, Gerald Penn, and
  Dong Yu. 2014.
\newblock Convolutional neural networks for speech recognition.
\newblock {\em IEEE/ACM Transactions on audio, speech, and language
  processing\/} 22(10):1533--1545.

\bibitem[{Amit and Geman(1997)}]{amit1997shape}
Yali Amit and Donald Geman. 1997.
\newblock Shape quantization and recognition with randomized trees.
\newblock {\em Neural computation\/} 9(7):1545--1588.

\bibitem[{Breiman(1996)}]{breiman1996bagging}
Leo Breiman. 1996.
\newblock Bagging predictors.
\newblock {\em Machine learning\/} 24(2):123--140.

\bibitem[{Breiman(2001)}]{breiman2001random}
Leo Breiman. 2001.
\newblock Random forests.
\newblock {\em Machine learning\/} 45(1):5--32.

\bibitem[{Cortes and Vapnik(1995)}]{cortes1995support}
Corinna Cortes and Vladimir Vapnik. 1995.
\newblock Support-vector networks.
\newblock {\em Machine learning\/} 20(3):273--297.

\bibitem[{Davis and Mermelstein(1990)}]{davis1990comparison}
Steven~B Davis and Paul Mermelstein. 1990.
\newblock Comparison of parametric representations for monosyllabic word
  recognition in continuously spoken sentences.
\newblock In {\em Readings in speech recognition\/}, Elsevier, pages 65--74.

\bibitem[{Ellis(2007)}]{ellis2007chroma}
Dan Ellis. 2007.
\newblock Chroma feature analysis and synthesis.
\newblock {\em Resources of Laboratory for the Recognition and Organization of
  Speech and Audio-LabROSA\/} .

\bibitem[{Friedman(2001)}]{friedman2001greedy}
Jerome~H Friedman. 2001.
\newblock Greedy function approximation: a gradient boosting machine.
\newblock {\em Annals of statistics\/} pages 1189--1232.

\bibitem[{Gemmeke et~al.(2017)Gemmeke, Ellis, Freedman, Jansen, Lawrence,
  Moore, Plakal, and Ritter}]{gemmeke2017audio}
Jort~F Gemmeke, Daniel~PW Ellis, Dylan Freedman, Aren Jansen, Wade Lawrence,
  R~Channing Moore, Manoj Plakal, and Marvin Ritter. 2017.
\newblock Audio set: An ontology and human-labeled dataset for audio events.
\newblock In {\em Acoustics, Speech and Signal Processing (ICASSP), 2017 IEEE
  International Conference on\/}. IEEE, pages 776--780.

\bibitem[{Gonzalez(2006)}]{gonzalez2006youtube}
Ricardo~Garcia Gonzalez. 2006.
\newblock Youtube-dl: download videos from youtube. com.

\bibitem[{Gouyon et~al.(2000)Gouyon, Pachet, Delerue et~al.}]{gouyon2000use}
Fabien Gouyon, Fran{\c{c}}ois Pachet, Olivier Delerue, et~al. 2000.
\newblock On the use of zero-crossing rate for an application of classification
  of percussive sounds.
\newblock In {\em Proceedings of the COST G-6 conference on Digital Audio
  Effects (DAFX-00), Verona, Italy\/}.

\bibitem[{Grosche et~al.(2010)Grosche, M{\"u}ller, and
  Kurth}]{grosche2010cyclic}
Peter Grosche, Meinard M{\"u}ller, and Frank Kurth. 2010.
\newblock Cyclic tempogram—a mid-level tempo representation for musicsignals.
\newblock In {\em Acoustics Speech and Signal Processing (ICASSP), 2010 IEEE
  International Conference on\/}. IEEE, pages 5522--5525.

\bibitem[{Hastie et~al.(2001)Hastie, Tibshirani, and
  Friedman}]{hastie2001elements}
Trevor Hastie, Robert Tibshirani, and Jerome Friedman. 2001.
\newblock The elements of statistical learnine.

\bibitem[{Ittichaichareon et~al.(2012)Ittichaichareon, Suksri, and
  Yingthawornsuk}]{ittichaichareon2012speech}
Chadawan Ittichaichareon, Siwat Suksri, and Thaweesak Yingthawornsuk. 2012.
\newblock Speech recognition using mfcc.
\newblock In {\em International Conference on Computer Graphics, Simulation and
  Modeling (ICGSM'2012) July\/}. pages 28--29.

\bibitem[{Jiang et~al.(2002)Jiang, Lu, Zhang, Tao, and Cai}]{jiang2002music}
Dan-Ning Jiang, Lie Lu, Hong-Jiang Zhang, Jian-Hua Tao, and Lian-Hong Cai.
  2002.
\newblock Music type classification by spectral contrast feature.
\newblock In {\em Multimedia and Expo, 2002. ICME'02. Proceedings. 2002 IEEE
  International Conference on\/}. IEEE, volume~1, pages 113--116.

\bibitem[{Kim and Stern(2012)}]{kim2012power}
Chanwoo Kim and Richard~M Stern. 2012.
\newblock Power-normalized cepstral coefficients (pncc) for robust speech
  recognition.
\newblock In {\em Acoustics, Speech and Signal Processing (ICASSP), 2012 IEEE
  International Conference on\/}. IEEE, pages 4101--4104.

\bibitem[{Kingma and Ba(2014)}]{kingma2014adam}
Diederik~P Kingma and Jimmy Ba. 2014.
\newblock Adam: A method for stochastic optimization.
\newblock {\em arXiv preprint arXiv:1412.6980\/} .

\bibitem[{Krizhevsky et~al.(2012)Krizhevsky, Sutskever, and
  Hinton}]{krizhevsky2012imagenet}
Alex Krizhevsky, Ilya Sutskever, and Geoffrey~E Hinton. 2012.
\newblock Imagenet classification with deep convolutional neural networks.
\newblock In {\em Advances in neural information processing systems\/}. pages
  1097--1105.

\bibitem[{Li et~al.(2010)Li, Chan, and Chun}]{li2010automatic}
Tom~LH Li, Antoni~B Chan, and A~Chun. 2010.
\newblock Automatic musical pattern feature extraction using convolutional
  neural network.
\newblock In {\em Proc. Int. Conf. Data Mining and Applications\/}.

\bibitem[{Lidy and Rauber(2005)}]{lidy2005evaluation}
Thomas Lidy and Andreas Rauber. 2005.
\newblock Evaluation of feature extractors and psycho-acoustic transformations
  for music genre classification.
\newblock In {\em ISMIR\/}. pages 34--41.

\bibitem[{Lidy and Schindler(2016)}]{lidy2016parallel}
Thomas Lidy and Alexander Schindler. 2016.
\newblock Parallel convolutional neural networks for music genre and mood
  classification.
\newblock {\em MIREX2016\/} .

\bibitem[{Mandel and Ellis(2005)}]{mandel2005song}
Michael~I Mandel and Dan Ellis. 2005.
\newblock Song-level features and support vector machines for music
  classification.
\newblock In {\em ISMIR\/}. volume 2005, pages 594--599.

\bibitem[{Nanni et~al.(2016)Nanni, Costa, Lumini, Kim, and
  Baek}]{nanni2016combining}
Loris Nanni, Yandre~MG Costa, Alessandra Lumini, Moo~Young Kim, and Seung~Ryul
  Baek. 2016.
\newblock Combining visual and acoustic features for music genre
  classification.
\newblock {\em Expert Systems with Applications\/} 45:108--117.

\bibitem[{Ng(2004)}]{ng2004feature}
Andrew~Y Ng. 2004.
\newblock Feature selection, l 1 vs. l 2 regularization, and rotational
  invariance.
\newblock In {\em Proceedings of the twenty-first international conference on
  Machine learning\/}. ACM, page~78.

\bibitem[{Scaringella and Zoia(2005)}]{scaringella2005modeling}
Nicolas Scaringella and Giorgio Zoia. 2005.
\newblock On the modeling of time information for automatic genre recognition
  systems in audio signals.
\newblock In {\em ISMIR\/}. pages 666--671.

\bibitem[{Simonyan and Zisserman(2014)}]{simonyan2014very}
Karen Simonyan and Andrew Zisserman. 2014.
\newblock Very deep convolutional networks for large-scale image recognition.
\newblock {\em arXiv preprint arXiv:1409.1556\/} .

\bibitem[{Soltau et~al.(1998)Soltau, Schultz, Westphal, and
  Waibel}]{soltau1998recognition}
Hagen Soltau, Tanja Schultz, Martin Westphal, and Alex Waibel. 1998.
\newblock Recognition of music types.
\newblock In {\em Acoustics, Speech and Signal Processing, 1998. Proceedings of
  the 1998 IEEE International Conference on\/}. IEEE, volume~2, pages
  1137--1140.

\bibitem[{Srivastava et~al.(2014)Srivastava, Hinton, Krizhevsky, Sutskever, and
  Salakhutdinov}]{srivastava2014dropout}
Nitish Srivastava, Geoffrey Hinton, Alex Krizhevsky, Ilya Sutskever, and Ruslan
  Salakhutdinov. 2014.
\newblock Dropout: A simple way to prevent neural networks from overfitting.
\newblock {\em The Journal of Machine Learning Research\/} 15(1):1929--1958.

\bibitem[{Tjoa(2017)}]{mir}
Steve Tjoa. 2017.
\newblock Music information retrieval.
\newblock \url{https://musicinformationretrieval.com/spectral_features.html}.
\newblock Accessed: 2018-02-20.

\bibitem[{Tomar(2006)}]{tomar2006converting}
Suramya Tomar. 2006.
\newblock Converting video formats with ffmpeg.
\newblock {\em Linux Journal\/} 2006(146):10.

\bibitem[{Tzanetakis and Cook(2002)}]{tzanetakis2002musical}
George Tzanetakis and Perry Cook. 2002.
\newblock Musical genre classification of audio signals.
\newblock {\em IEEE Transactions on speech and audio processing\/}
  10(5):293--302.

\bibitem[{Van Den~Oord et~al.(2016)Van Den~Oord, Dieleman, Zen, Simonyan,
  Vinyals, Graves, Kalchbrenner, Senior, and Kavukcuoglu}]{van2016wavenet}
Aaron Van Den~Oord, Sander Dieleman, Heiga Zen, Karen Simonyan, Oriol Vinyals,
  Alex Graves, Nal Kalchbrenner, Andrew Senior, and Koray Kavukcuoglu. 2016.
\newblock Wavenet: A generative model for raw audio.
\newblock {\em arXiv preprint arXiv:1609.03499\/} .

\bibitem[{Wyse(2017)}]{wyse2017audio}
Lonce Wyse. 2017.
\newblock Audio spectrogram representations for processing with convolutional
  neural networks.
\newblock {\em arXiv preprint arXiv:1706.09559\/} .

\bibitem[{Zwicker and Fastl(1999)}]{zwicker1999psychoacoustics}
E~Zwicker and H~Fastl. 1999.
\newblock Psychoacoustics facts and models .

\end{thebibliography}
\bibliographystyle{acl_natbib}
\end{document}